\documentclass{article}

\usepackage{arxiv}
\usepackage{setspace}
\usepackage[utf8]{inputenc} % allow utf-8 input
\usepackage[T1]{fontenc}    % use 8-bit T1 fonts
\usepackage{hyperref}       % hyperlinks
\usepackage{url}            % simple URL typesetting
\usepackage{booktabs}       % professional-quality tables
\usepackage{amsfonts}       % blackboard math symbols
\usepackage{nicefrac}       % compact symbols for 1/2, etc.
\usepackage{microtype}      % microtypography
\usepackage{lipsum}		% Can be removed after putting your text content
\usepackage{graphicx}
\usepackage{amsmath}
\usepackage[a]{esvect}

\title{Discrete Variable Representation method in the study of few-body quantum systems with non-zero angular momentum}

\author{\hspace{1mm}Vladimir A.~Timoshenko \\
	Saint-Petersburg State University\\
	\texttt{vladimir.timoshenko7@gmail.com} \\
	\And
	Evgeny A.~Yarevsky \\
	Saint-Petersburg State University\\
}

\date{}

\renewcommand{\headeright}{}
\renewcommand{\undertitle}{}

\hypersetup{
pdftitle={Discrete Variable Representation method in the study of few-body quantum systems with non-zero angular momentum},
pdfsubject={},
pdfauthor={Vladimir A.~Timoshenko, Evgeny A.~Yarevsky},
pdfkeywords={Few-body systems, energy spectrum, discrete-variable representation},
}
\begin{document}
\maketitle

\begin{abstract}
The systems with small binding energies and widely distributed in space bound-state wave functions are considered. 
Because the interaction potential is weak and rather localized compared to the characteristic sizes of wave functions of these systems, the problem of an accurate determination of binding energy and wave functions is complicated. 
%Slight adjustment in input parameters or an inaccuracy of calculations can lead to significant deviations of calculating results from true values. 
An essential part of the study is the development and application of the discrete-variable representation (DVR) method. 
This method is based on the determination of basis functions and the nodes and weights of a quadrature formula in such way that the values of a function are zero at all these nodes but one. 
With this representation the time required for calculating the Hamiltonian matrix elements is substantially reduced. 
The binding energies of several systems consisting of helium and lithium atoms have been obtained using the DVR method. 
%Thanks to the application of this approach, the calculation time was significantly reduced without loss of accuracy.

\end{abstract}

% keywords can be removed

\section{Introduction}

Systems of particles with small binding energies and wave functions that are widely distributed in space are considered in this work. 
The study of quantum-mechanical systems consisting of few particles can be a complicated problem. 
This is true for weakly bound systems, e.g. systems consisting
of helium and lithium atoms. 
The research of these systems is not an easy task and requires different approaches, solution methods and additional computational recourses. 
Extensive studies of weakly bound systems have been performed, see papers \cite{Esry, Motovilov_binding_energies, Gonzalez} and references therein.
 
The goal of this work is the further development and implementation of the discrete variables representation (DVR) method \cite{Baye,Shizgal,Light}.
This method allows to carry out calculations with smaller computing resources without loss of accuracy, and to reduce the calculation time. 
Due to the properties of the DVR functions, the calculation of the matrix elements of the potential energy can be substantially simplified. 
This simplification allows to increase the efficiency and, therefore, the accuracy of the solution.

Originally, our method was developed and implemented for quantum systems with zero orbital momentum.
The DVR decomposition was performed according to basis functions constructed on the base of the Legendre polynomials. 
In this paper, the algorithm has been generalised to perform calculations for systems with non-zero total orbital momentum. 
In order to make calculations faster, the DVR functions based on the Jacobi and associated Legendre polynomials are investigated.

The developed algorithm was used to calculate the binding energies of the systems Ne$_3$, He$_3$, Li--He$_2$ with the non-zero angular momentum. 
The binding energies of the Li-He$_2$ systems are calculated and compared with
theoretical results of other authors \cite{Yuan, Baccarelli, Kolganova_LiHe2}.

\section{Discrete-variable representation for the non-zero angular momentum}
\label{sec:theory}

Jacobi coordinates have been chosen for the three-body systems. Variable $x$ is the distance between particles 2 and 3, $y$ is the distance between particle 1 and the center of mass of pair (23) and $\theta$ is the angle between $x$ and $y$. Using an expansion of the wave function in terms of Wigner D-function, one can derive the  Hamiltonian of a three-particle system for states with non-zero angular momentum $J$, and its projection $M$, wich is composed of the diagonal and off-diagonal blocks~\cite{Yarevsky_thesis}
\begin{equation}
H^J_{MM}= -\dfrac{1}{x}\dfrac{\partial^2}{\partial x^2}-\dfrac{1}{y}\dfrac{\partial^2}{\partial y^2}y+\dfrac{J(J+1)-2M^2}{y^2}+V(x,y,\theta)-
\left(\dfrac{1}{x^2}+\dfrac{1}{y^2}\right)\left( \dfrac{\partial^2}{\partial\theta^2}+\cot\theta\dfrac{\partial}{\partial\theta}-\dfrac{M^2}{\sin^2\theta} \right),
\end{equation}
\begin{equation}
H^J_{MM'}= H^J_{MM'}\delta_{M,M\pm1} = \pm\dfrac{\lambda_{\pm}(J,M)}{y^2}\sqrt{1+\delta_{M0}\delta_{M'1}+\delta_{M1}\delta_{M'0}}\times\left(\dfrac{\partial}{\partial\theta}+(1\pm M)\cot\theta\right).
\end{equation}
Here, the potential $V = V(x, y, \theta)$ is a sum of two-particle potentials that depend on inter-particle distances only, and $\lambda_{\pm}(J,M)=\sqrt{J(J+1)-M(M\pm1)}$.

Let us rewrite the kinetic-energy operator in terms of $z = \cos \theta$ and apply the DVR approach~\cite{Light} for the variable z. 
For the angular parts of the diagonal and off-diagonal components of the kinetic energy operator we have
\begin{equation}
\dfrac{d^2}{d\theta^2}+\cot\theta\dfrac{d}{d\theta}-\dfrac{M^2}{\sin^2\theta}=(1-z^2)\dfrac{d^2}{dz^2}-2z\dfrac{d}{dz}-\dfrac{M^2}{1-z^2},
\end{equation}
\begin{equation}
\dfrac{d}{d\theta}+(1\pm M)\cot\theta=-\sqrt{1-z^2}\dfrac{d}{dz}+(1\pm M)\dfrac{z}{\sqrt{1-z^2}}.
\end{equation}

The DVR functions $\varphi_i (z)$ and their derivatives $\varphi^\prime_i(z)$ are constructed with the orthogonal polynomials $P_n(z)$ and are related to the Gauss-type quadrature formulas. 
The following property of the DVR-functions allows us to simplify the calculations of the potential energy operator:
\begin{equation}
  \varphi_i(z) = \dfrac{P_n(z)}{P’_n(z_i)(z-z_i)},\hspace{1 cm}
 	\varphi_i(z_k)=\delta_{ik}.
 \end{equation}
Here, $z_1,\dots,z_n$ are the zeros of the polynomial $P_n(z)$.

In order to obtain the matrix of the kinetic energy operator, it is necessary to get the first derivatives $\varphi'_i(z)$ at the points $z_1,\dots,z_n$. 
Substituting the $P_n(z)$ with its Taylor series at the points $z_i$, we get for the values of the derivatives $\varphi’_i(z_i)$:
%Considering $P_n(z_k)=0$
 \begin{equation}
  \label{ddvr-function2}
 	\varphi'_i(z_k)= \dfrac{P’_n(z_k)}{P’_n(z_i)(z_k-z_i)}\: (\text{for }k \neq i),\hspace{1 cm}
 	\varphi^{\prime}_i(z_i)=-\dfrac{P^{\prime \prime}_{n}(z_i)}{2P^{\prime}_{n}(z_i)}.
 \end{equation}

\subsection{Jacobi polynomials}

An integral over $[-1,\;1]$ can be approximated with the Gaussian quadrature  rule~\cite{Abramovitz} based on the Jacobi polynomials $P^{(\alpha, \beta)}(z)$:
 \begin{equation}
  \int_{-1}^1 f(z)\,dz \approx \sum_{i=1}^n \dfrac{w_i}{\rho(z_i)}f(z_i),
 \end{equation}
where $\rho(z) = (1-z)^\alpha(1+z)^\beta$ is a weight function, $z_1,\dots,z_n$ are the roots of the polynomial $P_n$, $P_n(z_i)=0$.
The weights $w_i$ can be evaluated explicitly,
 \begin{equation}
 	w_i=-\dfrac{2n+\alpha+\beta+2}{n+\alpha+\beta+1}\dfrac{\Gamma(n+\alpha+1)\Gamma(n+\beta+1)}{\Gamma(n+\alpha+\beta+1)(n+1)!}\dfrac{2^{\alpha+\beta}}{P^{(\alpha, \beta)}_n{}'(z_i)P^{(\alpha, \beta)}_{n+1}(z_i)}.
 \end{equation}
As was mentioned above, the potential energy operator in the DVR representation is simplified, and is diagonal
  \begin{eqnarray}
 	  V_{ij}= \int_{-1}^1 \dfrac{\varphi_i(z)}{\sqrt{w_i}}V(x,y,z)\frac{\varphi_j(z)}{\sqrt{w_j}} dz \approx \sum_k \dfrac{w_k}{\rho(z_k)} V(x,y,z_k)\dfrac{\varphi_i(z_k)}{\sqrt{w_i}}\frac{\varphi_j(z_k)}{\sqrt{w_j}}=
	  \dfrac{V(x,y,z_i)}{\rho(z_i)}\delta_{ij}.
 \end{eqnarray}  
Due to the properties of the Jacobi polynomials, the derivative $\varphi’_i(z_i)$ can be evaluated explicitly:
 \begin{equation}
 	\varphi'_i(z_i)=\dfrac{\beta-\alpha-(\alpha+\beta+2)z_i}{2(1-z_i^2)}.
 \end{equation}

\subsection{Associated Legendre polynomials}
Let us now construct the DVR-functions based on the associated Legendre polynomials. 
There is an expectation that the factor $\sin^m\theta$ in the associated Legendre polynomials $P^m_{n+m}(z=\cos\theta)$ used as a weight function will provide more accurate calculations.
The associated Legendre polynomials can be expressed as~\cite{Abramovitz}:
\begin{equation}
P^m_{n+m}(z) = (-1)^m \dfrac{(n+2m)!}{2^m (n+m)!} \cdot(1-z^2)^{m/2} P^{(m,m)}_n(z).
\end{equation}
The zeros $z_1,\dots,z_n$ of the polynomial $P^m_{n+m}(z)$ different from $\pm 1$ coincide with the zeros of $P^{(m,m)}_n(z)$. 
Thus, the DVR-function constructed with the associated Legendre polynomials are expressed as:
\begin{equation}
   \varphi_i(z) = \dfrac{P^m_{n+m}(z)}{P^{m \prime}_{n+m}(z-z_i)}=\left(\dfrac{1-z^2}{1-z^2_i}\right) ^{m/2} \cdot \dfrac{P^{(m,m)}_n(z)}{P^{(m,m) \prime}_n(z_i)(z-z_i)},
 \hspace{1 cm}
 	\varphi_i(z_k)=\delta_{ik} .
\end{equation}
Using the equations for the associated Legendre polynomials, we can express  
$P^{m \prime \prime}_{n+m}(z)$ in terms of $P^{m \prime}_{n+m}(z)$ at the roots $z_i$:
 \begin{equation}\label{ddJ-polynomial}
 	  P^{m \prime \prime}_{n+m}(z_i)=\dfrac{2z_i P^{m \prime}_{n+m}(z_i)}{1-z_i^2}.
 \end{equation}
Substituting (\ref{ddJ-polynomial}) into (\ref{ddvr-function2}), we finally obtain:
  \begin{equation}
 	\varphi^{\prime}_i(z_i)=-\dfrac{z_i}{1-z_i^2}.
 \end{equation}

\section{Results and descation}
\label{sec:results}
The approach combining the finite-element method~\cite{neon} for the coordinates $x$ and $y$, and the DVR method for the coordinate $z$ has been developed for calculating the binding energies of the three particle quantum systems.
The energy levels of weakly bound systems ${}^6$Li--He${}_2$ and ${}^7$Li--He${}_2$ have been calculated with the DVR method based on the Legendre polynomials, $\alpha=\beta=0$. 
Due to the use of the DVR, the calculation time has been significantly reduced without loss of accuracy.
The results for binding energies, computation times and relative inaccuracies $\delta E = |{E}/{(E_{\text{exact}}-E)}|$ for the ${}^7$Li--He${}_2$ system are shown in Table~1.
One can see that the results of calculations without the DVR demonstrate the variational behaviour while the DVR results approach the exact value from below.
Both results converge to the same exact value when the accuracy increases.

\begin{table}[h!]
\begin{center}
\caption{The binding energies $E_1$, computation times $t$, and relative errors $\delta E_1$ for the different number $n$ of functions in the DVR expansion. 
	The results for the ${}^7$Li--He${}_2$ system are shown.}
\label{tab:LiHe2DVR}
\begin{tabular}{|c|c|c|c|c|c|} 
\hline
  n & 5 & 10 & 15 & 20 & 25  \\ \hline
  \multicolumn{6}{|c|}{Legendre polynomial expansion without the DVR} \\ \hline
  $E_1$, cm${}^{-1}$ & -3.44$\cdot 10^{-7}$  & -3.00$\cdot 10^{-2}$    & -3.94$\cdot 10^{-2}$   & -4.07$\cdot 10^{-2}$ & -4.12$\cdot 10^{-2}$ \\ \hline
  $\delta E_1$, cm${}^{-1}$ &   1.00 & 2.75$\cdot 10^{-1}$    & 4.67$\cdot 10^{-2}$   & 1.49$\cdot 10^{-2}$ & 2.26$\cdot 10^{-3}$ \\ \hline
  $t$, sec  & 6.0 & 36.6 &  119.6 & 287.9 & 553.5  \\ \hline
  \multicolumn{6}{|c|}{Legendre polynomial expansion with the DVR} \\ \hline
  $E_1$, cm${}^{-1}$ & -4.81$\cdot 10^{-2}$  & -4.67$\cdot 10^{-2}$    & -4.19$\cdot 10^{-2}$   & -4.15$\cdot 10^{-2}$ & -4.14$\cdot 10^{-2}$ \\ \hline
  $\delta E_1$, cm${}^{-1}$ &   1.63$\cdot 10^{-1}$ & 1.31$\cdot 10^{-1}$    & 1.41$\cdot 10^{-2}$   & 5.10$\cdot 10^{-3}$ & 1.96$\cdot 10^{-3}$ \\ \hline
  $t$, sec & 2.2 & 8.2 &  19.1 & 35.2 & 65.8  \\ \hline
%    \multicolumn{6}{|c|}{time reduction factor, $t_\text{wo DVR}/t_\text{with DVR}$ } \\ \hline
%  & 2.7 & 4.5 & 6.3 & 8.2 & 8.4 \\ \hline 
  \end{tabular}
\end{center}
\end{table}

Calculated binding energies for the ${}^6$Li--He${}_2$ and  ${}^7$Li--He${}_2$ systems are shown in Table~2 together with the comparison to the theoretical results of other authors~\cite{Yuan,Baccarelli,Kolganova_LiHe2}.
For the TTY+KTTY and LM2M2+KTTY potentials, our binding energies are considerably deeper. 
As our results are almost variational, we believe that they are closer to the exact values.

\begin{table}[h!]
\begin{center}
\caption{Binding energies of the Li--He${}_2$ system (cm${}^{-1}$) for different interparticle potentials: TTY~\cite{TTY}, LM2M2~\cite{LM2M2}, Cvetko~\cite{Cvetko}.}
\label{tab:LiHe2}
\begin{tabular}{|c|c|c|c|c|}
\hline
  & He-He potential& Li-He potential & ${}^6$Li--He${}_2$ & ${}^7$Li--He${}_2$ \\ \hline
 paper~\cite{Yuan}& TTY & KTTY &-2.18$\cdot 10^{-2}$ & -3.18$\cdot 10^{-2}$ \\ \hline
 This work & TTY & KTTY &-3.71$\cdot 10^{-2}$&-5.41$\cdot 10^{-2}$ \\ \hline
 paper~\cite{Baccarelli}& LM2M2 & Cvetko &-3.61$\cdot 10^{-2}$ & -5.10$\cdot10^{-2}$ \\ \hline
 This work  & LM2M2& Cvetko & -2.62$\cdot 10^{-2}$ & -4.07$\cdot 10^{-2}$ \\ \hline
 paper~\cite{Kolganova_LiHe2} & LM2M2 & KTTY & -2.46$\cdot 10^{-2}$ & -3.54$\cdot 10^{-2}$ \\ \hline
 This work & LM2M2 & KTTY & -3.71$\cdot 10^{-2}$&-5.41$\cdot 10^{-2}$ \\ \hline
 \end{tabular}
\end{center}
\end{table}

The DVR approach has also been implemented with the DVR functions constructed with the Jacobi polynomials $P_n^{(\alpha,\beta)}(z)$ and the associated Legendre polynomials $P^m_{n+m}(z)$. 
Using of these polynomials make it possible to choose parameters $\alpha$, $\beta$ and $m$ such that the weight $\rho(z)$ countervails the interaction potential.
This approach has been used for calculating the binding energy of the neon trimer.
The results for different number $n$ of the functions in the representation are presented in Table~3. 
Here the total angular momentum $J=1$, and the energy levels correspond to the positive symmetry.
 \begin{table}[t]
\begin{center}
\caption{Binding energies (in cm${}^{-1}$) of the Ne${}_3$ for different DVR-functions based on Jacobi polynomials, $\alpha=\beta=0,-0.5$, and the associated Legendre polynomials, $\alpha=\beta=m$.}
\label{tab:compareAL}
\begin{tabular}{|c|c|c|c|c|}
\hline
  & $n=5$ & $n=10$ & $n=15$ & $n=20$ \\ \hline
  $\alpha=\beta$ & \multicolumn{4}{|c|}{$E_0$} \\ \hline
 -0.5 & -46.959 & -50.388 & -51.254 & -51.143 \\ \hline
    0 & -56.501 & -50.302 & -50.936 & -50.982 \\ \hline
    m & -56.484 & -56.484 & -50.930 & -50.980 \\ \hline

 $\alpha=\beta$ & \multicolumn{4}{|c|}{$E_1$} \\ \hline
 -0.5 & -46.827 & -50.386 & -50.386 & -51.138 \\ \hline
    0 & -56.483 & -50.297 & -50.936 & -50.981 \\ \hline
    m & -48.436 & -49.184 & -50.655 & -50.912 \\ \hline
    
    $\alpha=\beta$  & \multicolumn{4}{|c|}{$E_2$} \\ \hline
 -0.5 & -42.679	& -41.564	& -38.666	& -38.488\\ \hline
    0 & -44.070	& -37.640	& -38.273	& -38.320 \\ \hline
    m & -43.956	& -37.622	& -38.249	& -38.300 \\ \hline
    
   $\alpha=\beta$   & \multicolumn{4}{|c|}{$E_3$} \\ \hline
 -0.5 & -42.313	& -41.558	& -38.658	& -38.461 \\ \hline
    0 & -43.956	& -37.578	& -38.258	& -38.316 \\ \hline
    m & -38.656	& -37.228	& -38.137	& -38.277 \\ \hline
    
    $\alpha=\beta$  & \multicolumn{4}{|c|}{$E_4$} \\ \hline
 -0.5 & -38.582	& -37.382	& -38.418	& -38.379 \\ \hline
    0 & -38.882	& -37.306	& -38.154	& -38.215 \\ \hline
    m & -36.732	& -36.026	& -37.844	& -38.166 \\ \hline
 \end{tabular}
\end{center}
\end{table}

\paragraph[h!]{Conclusion.}
The method combining the finite-element approach and the DVR method was developed for calculating quantum three-particle systems. 
The energy levels of weakly bound systems consisting of several atoms were calculated.
Due to the application of the DVR method, the calculation time was significantly reduced without loss in accuracy.

The approach was extended by using various types of quadrature formulas to construct DVR functions, in particular, formulas based on Jacobi and associated Legendre polynomials. 
Such formulas allow us the specific features of pairwise potentials at short distances to be taken into account more accurately.

The next step of the research is the generalization of the proposed method to the 
complex-valued functions and will allow one to enhance the efficiency of searching for resonance states and the study of scattering processes.

\paragraph[h!]{Acknowledgements.}
The reported study was funded by RFBR, project number 19-32-90148.
The calculations were performed on resources of the Computational Center of St. Petersburg State University.

\bibliographystyle{unsrt}


\end{document}